\documentclass[conference]{IEEEtran}
\IEEEoverridecommandlockouts
\RequirePackage[utf8]{inputenc}
\RequirePackage[english]{babel}
\RequirePackage{cite}
\RequirePackage{amsmath,amssymb,amsfonts}
\RequirePackage{algorithmic}
\RequirePackage[pscoord]{eso-pic}
\RequirePackage{textcomp}
\usepackage{acronym}
\RequirePackage{xcolor}
\RequirePackage[a4paper, total={184mm,239mm}]{geometry}
\PassOptionsToPackage{detect-all, per-mode=symbol, range-units=single, range-phrase=~to~}{siunitx}
\RequirePackage{siunitx}
\RequirePackage{nicefrac}
\RequirePackage[hidelinks]{hyperref}
\RequirePackage{orcidlink}
\RequirePackage[noabbrev,capitalise]{cleveref}
\RequirePackage{float}
\RequirePackage[caption=false]{subfig}
\RequirePackage{url}
\RequirePackage{lipsum}
\RequirePackage{placeins}
\RequirePackage{xfrac}

\RequirePackage{amssymb}
\RequirePackage{pifont}

\RequirePackage{threeparttable}
\RequirePackage{booktabs}
\RequirePackage{colortbl}
\RequirePackage{tabularx}
\RequirePackage{makecell}

\RequirePackage{graphicx}
\RequirePackage{glossaries}
\RequirePackage{xspace}
\usepackage{todonotes}
\usepackage[font=small, skip=3pt]{caption} %

\begin{document}

\AddToShipoutPictureBG*{%
  \AtPageUpperLeft{%
    \hspace{\paperwidth}%
    \raisebox{-\baselineskip}{%
      \makebox[-35pt][r]{\footnotesize{
        \copyright~2025~IEEE. Personal use of this material is permitted. %
        Permission from IEEE must be obtained for all other uses, in any current or future media, including
      }}
}}}%

\AddToShipoutPictureBG*{%
  \AtPageUpperLeft{%
    \hspace{\paperwidth}%
   \raisebox{-2\baselineskip}{%
      \makebox[-37pt][r]{\footnotesize{
        reprinting/republishing this material for advertising or promotional purposes, creating new collective orks, for resale or redistribution to servers or lists, or
      }}
}}}%

\AddToShipoutPictureBG*{%
  \AtPageUpperLeft{%
    \hspace{\paperwidth}%
    \raisebox{-3\baselineskip}{%
      \makebox[-185pt][r]{\footnotesize{
       reuse of any copyrighted component of this work in other works.
      }}
}}}%

\title{Towards Reliable Systems: A Scalable Approach to AXI4 Transaction Monitoring\\
}

\ifx\blind\undefined
    \author{
        \IEEEauthorblockN{%
        Chaoqun Liang\orcidlink{0009-0008-0556-7758}\IEEEauthorrefmark{1}, %
        Thomas Benz\orcidlink{0000-0002-0326-9676}\IEEEauthorrefmark{2}, %
        Alessandro Ottaviano\orcidlink{0009-0000-9924-3536}\IEEEauthorrefmark{2}, \\%
        Angelo Garofalo\orcidlink{0000-0002-7495-6895}\IEEEauthorrefmark{1}\IEEEauthorrefmark{2}, %
        Luca Benini\orcidlink{0000-0001-8068-3806}\IEEEauthorrefmark{1}\IEEEauthorrefmark{2}, %
        Davide Rossi\orcidlink{0000-0002-0651-5393}\IEEEauthorrefmark{1}%
        }
        \IEEEauthorblockA{
            \IEEEauthorrefmark{1}~\textit{Department of Electrical, Electronic, and Information Engineering, University of Bologna}, Italy \\
            \IEEEauthorrefmark{2}~\textit{Integrated Systems Laboratory, ETH Zurich}, Switzerland \\
        }
    }
\else
    \author{%
            \vspace{1.1cm} %
            \textit{Authors omitted for blind review}
            \vspace{1.1cm} %
            }
\fi

\maketitle

\begin{abstract}
In safety-critical SoC applications such as automotive and aerospace, reliable transaction monitoring is crucial for maintaining system integrity. This paper introduces a drop-in Transaction Monitoring Unit (TMU) for AXI4 subordinate endpoints that detects transaction failures including protocol violations or timeouts and triggers recovery by resetting the affected subordinates.

Two TMU variants address different constraints:  a Tiny-Counter solution for tightly area-constrained systems and a Full-Counter solution for critical subordinates in mixed-criticality SoCs. The Tiny-Counter employs a single counter per outstanding transaction, while the Full-Counter uses multiple counters to track distinct transaction stages, offering finer-grained monitoring and reducing detection latencies by up to hundreds of cycles at roughly 2.5× the area cost. The Full-Counter also provides detailed error logs for performance and bottleneck analysis.

Evaluations at both IP and system levels confirm the TMU’s effectiveness and low overhead. In GF12 technology, monitoring 16–32 outstanding transactions occupies 1330–2616\,µm\textsuperscript{2} for the Tiny-Counter and 3452–6787\,µm\textsuperscript{2} for the Full-Counter; moderate prescaler steps reduce these figures by 18–39\% and 19–32\%, respectively, with no loss of functionality. Results from a full-system integration demonstrate the TMU’s robust and precise monitoring capabilities in safety-critical SoC environments.
\end{abstract}


\section{Introduction}
Automotive and aerospace systems demand high reliability and effective fault management in system-on-chip (SoC) designs due to their safety-critical nature. These applications perform complex tasks that require high-performance processing units, advanced safety mechanisms, and robust fault tolerance and recovery strategies to ensure dependable operation under varying conditions. Any failure in the underlying integrated circuits (ICs) within these systems can lead to severe consequences, ranging from equipment malfunctions to complete system outages.

To achieve near-zero Defective Parts Per Million (DPPM) and comply with standards such as ISO 26262 and Automotive Safety Integrity Levels (ASILs) \cite{b0}, SoC designers must adopt stringent design guidelines that address potential faults and strengthen system resilience. Central to these guidelines is the reliability of the SoC’s communication bus, which orchestrates data exchanges between processors, memories, and peripheral devices. A dependable bus prevents errors such as data corruption or protocol violations from propagating, ensuring that high-bandwidth traffic is handled accurately and within required time constraints.
 
Among the most widely adopted bus protocols for SoCs is the Advanced eXtensible Interface 4 (AXI4) \cite{b1}, part of the ARM AMBA family. AXI4’s support for multiple outstanding transactions increases throughput and lowers latency, making it a strong candidate for complex, high-performance designs. However, this same flexibility introduces complexity that can undermine reliability if not properly safeguarded. Failures can originate from either the manager (e.g., host processor) or subordinate (e.g., memory controller, peripheral device). A subordinate device may fail to respond on time or return incorrect data, leading to deadlocks or data corruption. Conversely, a manager failure can stall the bus entirely, causing missed deadlines and incomplete critical tasks.

To address these challenges, this paper presents a Transaction Monitoring Unit (TMU) IP block for AXI4 interconnect endpoints. Positioned between each subordinate and the AXI4 bus, the TMU detects protocol violations and device timeouts in real-time, triggering fault recovery actions such as issuing a hardware reset to restore malfunctioning subordinates. By isolating faults quickly, the TMU ensures minimal disruption and continued system integrity.

The key contributions of the paper are as follows:
\begin{itemize}
\item \textbf{A modular Transaction Monitoring Unit (TMU):} 
Addresses two critical gaps in the state-of-the-art: scalability to monitor multiple outstanding transactions and precision in fault detection.

\item \textbf{Tiny-Counter solution:} Tracks 16 to 32 outstanding transactions within a compact area ranging from 1330 µm² to 2616 µm², suitable for monitoring typical subordinate devices within heterogeneous, high-performance automotive and aerospace SoCs.

\item \textbf{Full-Counter solution:} Monitors transactions at the stage level, detecting faults with a latency of just one clock cycle. It also logs performance metrics such as latency and throughput, ideal for critical subordinates requiring detailed analysis.

\item \textbf{Comprehensive design space exploration:} Demonstrates the TMU’s scalability in 12\,nm CMOS under four configurations, highlighting the trade-offs between area, monitoring granularity, and performance. Showcases the system's adaptability across different SoC designs.
    
\end{itemize}

\section{Architecture}

\begin{figure}[htbp]
    \centering
    \includegraphics[width=0.53\textwidth]{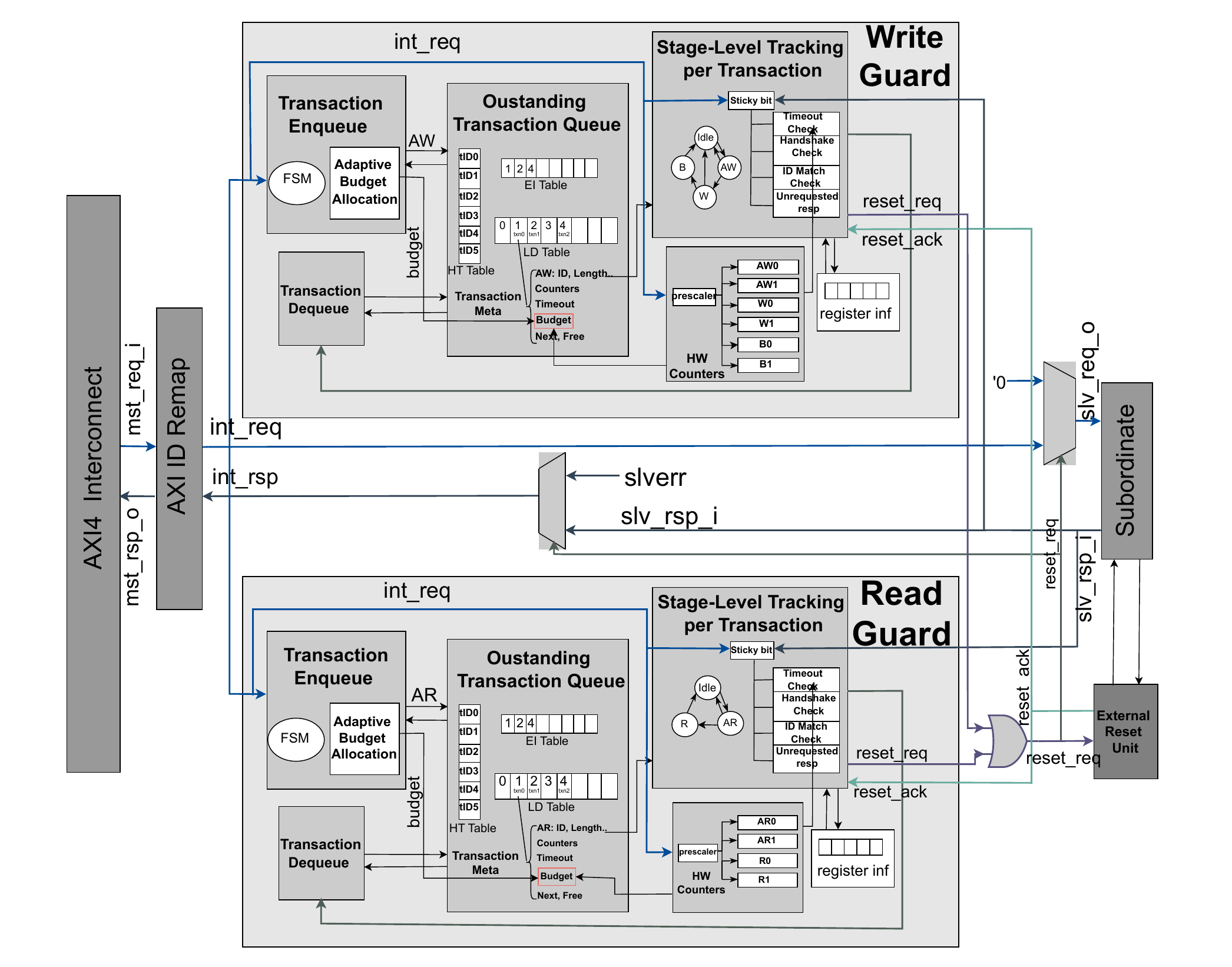}
    \caption{TMU Full-Counter (Fc) Architecture. Multiple counters track distinct phases of each transaction.}
    \label{fig:full_top}
\end{figure}

\begin{figure}[htbp]
    \centering
    \includegraphics[width=0.53\textwidth]{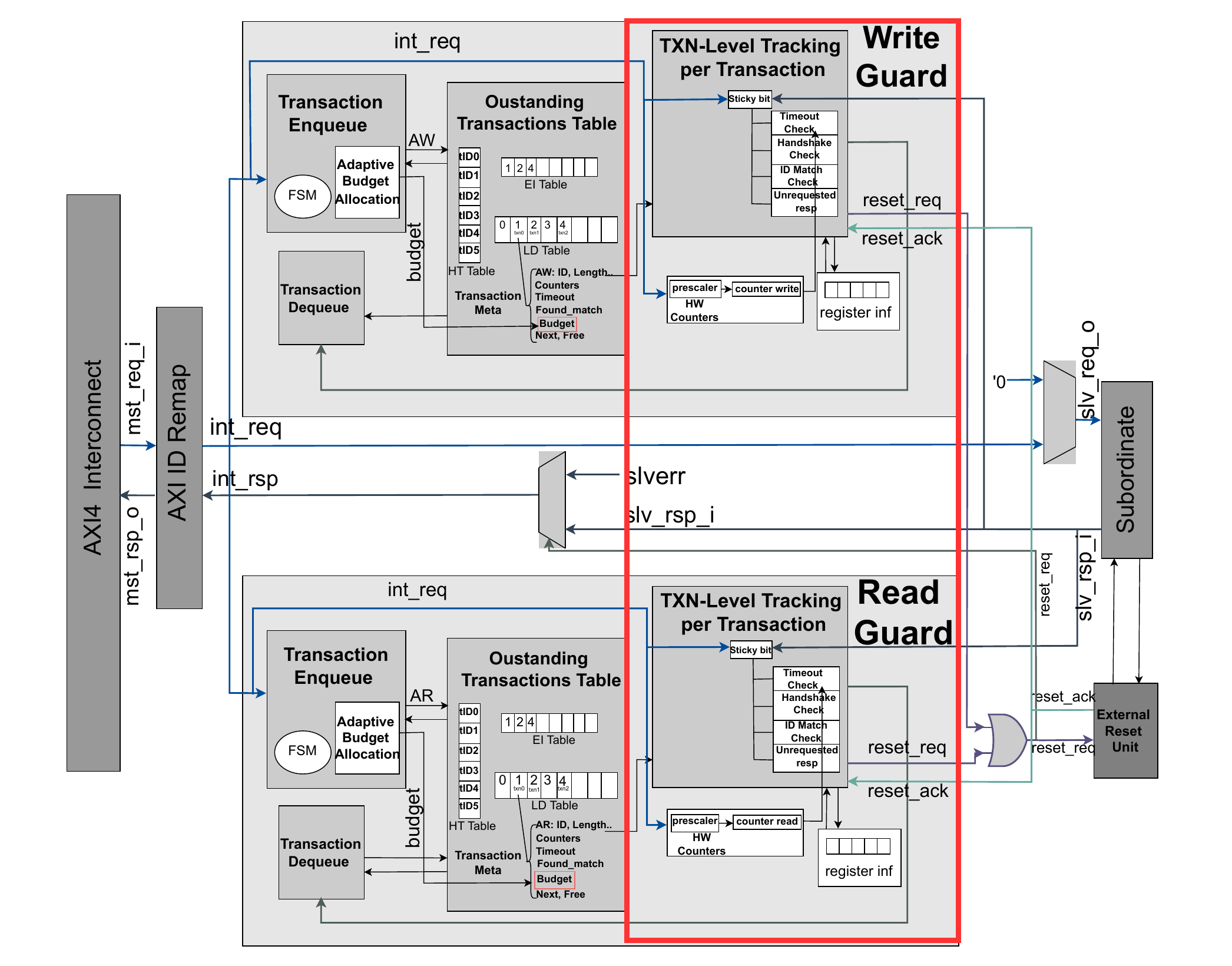}
    \caption{TMU Tiny-Counter (Tc) Architecture. Single counter tracks a transaction. The \textit{red square} highlights logic that differs in the Full-Counter.}
    \label{fig:tiny_top}
\end{figure}

The TMU resides between the AXI4 interconnect (manager side) and the subordinate device, continuously monitoring transactions for protocol violations or timeouts. It is available in two main variants: a \emph{Tiny-Counter (Tc)} for minimal area overhead, and a \emph{Full-Counter (Fc)} for more granular, phase-level analysis (Figs.~\ref{fig:full_top} and \ref{fig:tiny_top}).

\subsection{ID Optimization and Guard Modules}
To handle AXI4’s multiple ID lines efficiently, the TMU includes an \textit{AXI ID Remapper} that compacts a wide, sparse ID space into a narrower one. This remapping streamlines transaction tracking. Since AXI4 separates write and read channels, the TMU provides dedicated \textit{Write Guard} and \textit{Read Guard} modules that independently check protocol signals for correctness and timing. A set of software-configurable registers enables or disables the TMU and adjusts parameters such as time budgets, latency statistics, interrupt behavior, and error logging.

\subsection{Normal Operation and Fault Detection}

Under normal operation, transactions traverse from the manager (AXI4 interconnect) to the subordinate device without added latency, while the TMU listens in parallel. On detecting a protocol violation or timeout, the TMU severs both request and response paths to prevent error propagation. It then modifies the \textit{slverr} response to the manager, aborts outstanding transactions, and signals an external hardware reset unit \cite{b3} to reinitialize the faulty subordinate. Concurrently, it raises an interrupt that prompts the processor to run software-based recovery routines.

\subsection{Outstanding Transaction Tracking}
To handle concurrent outstanding transactions from multiple IDs, the TMU enqueues new write or read requests (\textit{aw\_valid} or \textit{ar\_valid}) into a 2D Outstanding Transaction Table (OTT).

\begin{figure}[htbp]
\centering
\includegraphics[width=0.5\textwidth]{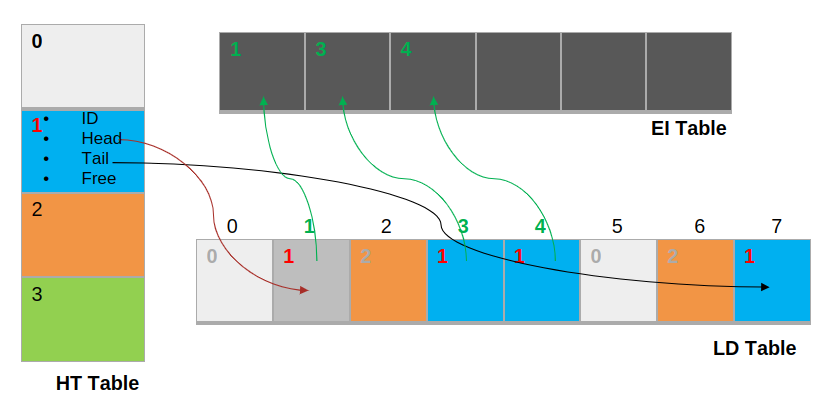}
\caption{Outstanding Transaction Table (OTT) architecture.}
\label{fig:oot}
\end{figure}

As shown in Fig.~\ref{fig:oot}, the OTT is divided into three linked subtables:
\begin{itemize}
    \item \textbf{ID Head-Tail (HT) Table}:  
    Maintains a FIFO structure for each \textit{tID} to ensure in-order completion of outstanding transactions sharing the same ID as required by AXI4. Each ID’s \textit{head} and \textit{tail} pointers link into the LD table.

    \item \textbf{Linked Data (LD) Table}:  
    Stores each outstanding transaction’s details, including its \textit{tID}, address, transaction state, budget, latency, timeout status.
    \item \textbf{Enqueue Index (EI) Table}:  
    Enforces the AXI4 requirement that write data on the W channel must follow the same order as the corresponding address transfers on the AW channel. By storing the sequence of AW/AR requests, the EI table ensures that each W beat is correctly associated and presented in the same order. Similarly, for reads, it aligns AR addresses with the R data phase though there is no strict ordering rules for read in AXI4.

\end{itemize}

\subsection{Tracking Capacity and Resource Management}
The TMU's capacity is determined by \textit{MaxUniqID} (the maximum number of unique IDs) and \textit{TxnPerUniqID} (the maximum outstanding transactions per ID). Together, these define \textit{MaxOutstdTxns} as shown in Table~\ref{tab:my-table}. When the OTT is saturated, new transaction requests are stalled until an existing transaction completes or is aborted to prevent overflows and preserve system integrity.

\begin{table}[ht]
\centering
\caption{Key Design Parameters}
\label{tab:my-table}
\resizebox{\columnwidth}{!}{%
\begin{tabular}{|l|l|}
\hline
\textbf{Parameter} & \textbf{Description} \\ \hline
\textbf{MaxUniqIDs} & Number of unique Transaction IDs that can be tracked \\ \hline
\textbf{TxnPerUniqID} & Outstanding transactions allowed per ID \\ \hline
\textbf{MaxOutstdTxns} & Total outstanding transactions supported \\ \hline
\end{tabular}
}
\end{table}

\subsection{Full-Counter vs. Tiny-Counter Solutions}

\textbf{Full-Counter (Fc)}: Allocates multiple counters per transaction to track distinct phases separately. For a write transaction, the TMU tracks six key phases ( Fig. 4): 
\begin{enumerate}
    \item \textbf{Address Handshake:} From \textit{aw\_valid} to \textit{aw\_ready}.
    \item \textbf{Data-Phase Entry:} From \textit{aw\_ready} to the first \textit{w\_valid}.
    \item \textbf{First Data Transfer Handshake:} From \textit{w\_valid} to \textit{w\_ready}.
    \item \textbf{Burst Data Transfer:} From \textit{w\_first} to \textit{w\_last}.
    \item \textbf{Response Monitoring:} From \textit{w\_last} to \textit{b\_valid}, including ID checks and correctness checks.
    \item \textbf{Response Readiness:} From \textit{b\_valid} to \textit{b\_ready}.
\end{enumerate}

As illustrated in Fig. 5, a read transaction similarly includes address, data, and response phases tracked by separate counters. Such phase-level timers allow earlier detection of stalls or protocol violations and detailed logging of performance metrics, at the cost of additional hardware.

\textbf{Tiny-Counter (Tc)}: The Tc variant illustrated in Fig. 6 assigns a single counter to each outstanding transaction, tracking it from initiation (\textit{aw\_valid}/\textit{ar\_valid}) until completion (\textit{b\_valid}/\textit{r\_last}). This approach minimizes hardware overhead but provides only transaction-level granularity: any fault is detected once the overall timeout expires.

\begin{figure}[htbp]
    \centering
    \includegraphics[width=0.45\textwidth,height=0.19\textheight]{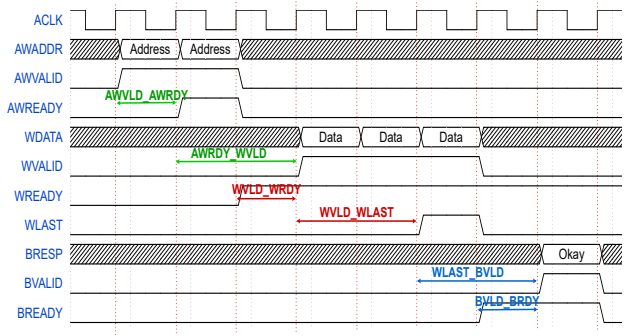}
    \caption{Phase-level monitoring of a write transaction in the Full-Counter (Fc) solution.}
    \label{fig:full_write}
\end{figure}

\begin{figure}[htbp]
    \centering
    \includegraphics[width=0.5\textwidth,height=0.15\textheight]{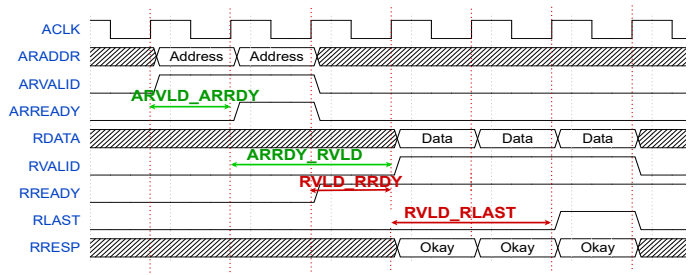}
    \caption{Phase-level monitoring of a read transaction in the Full-Counter (Fc) solution.}
    \label{fig:full_read}
\end{figure}

\begin{figure}[htbp]
    \centering
    \includegraphics[width=0.51\textwidth,height=0.19\textheight]{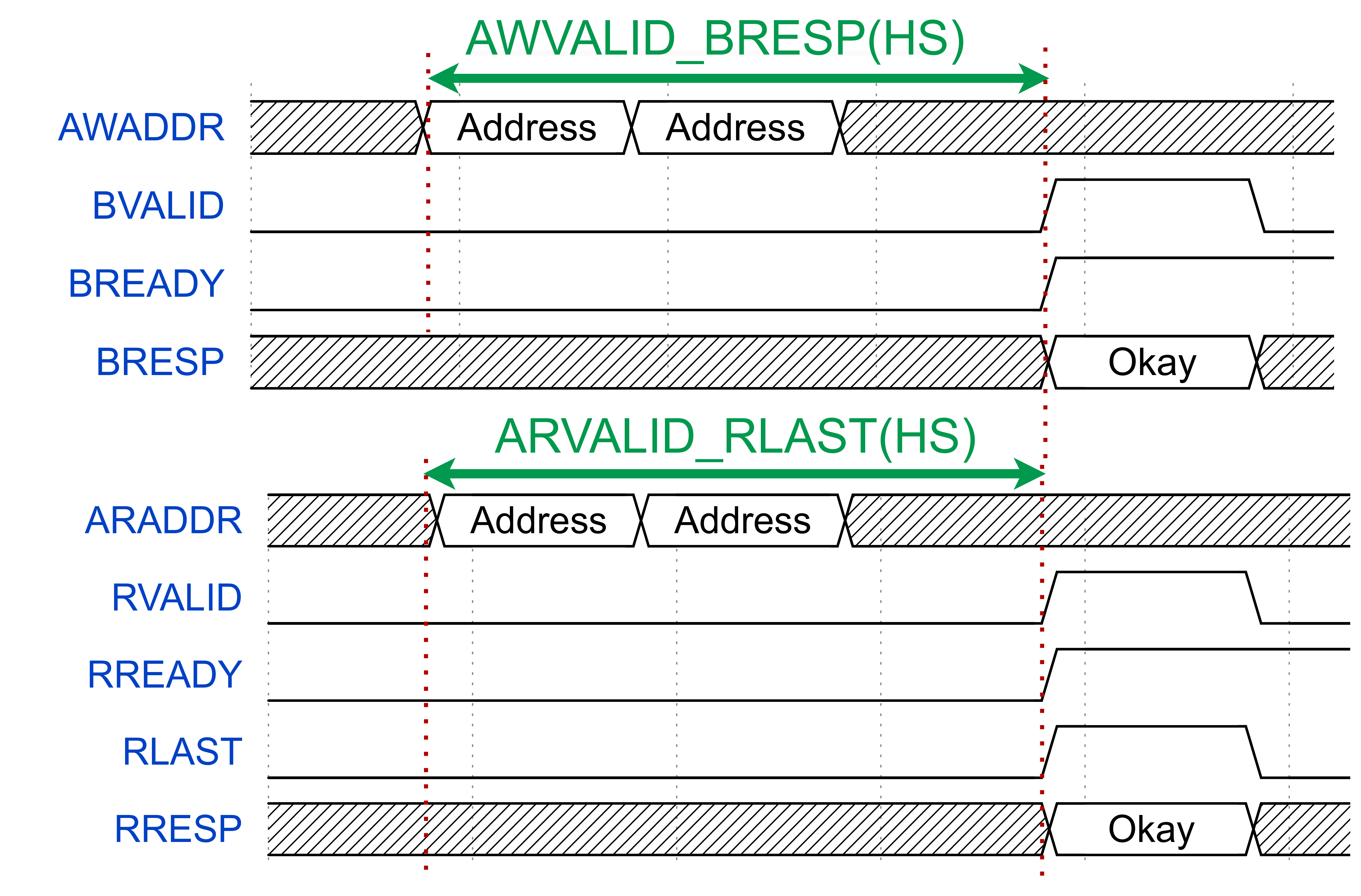}
    \caption{Transaction level monitoring in the Tiny-Counter (Tc) solution.}
    \label{fig:tiny_stage}
\end{figure}

\subsection{Adaptive Time-Budgeting Mechanism}

To avoid false timeouts in systems with large bursts or burst chaining, the TMU adapts its time budgets based on both burst length and accumulated outstanding traffic in the OTT. For both Fc and Tc solutions, these adaptive budgets are typically divided into two components:
\begin{itemize}
    \item \textbf{Queue Waiting Time:}  
    From the address handshake (\textit{aw\_valid} or \textit{ar\_valid}) to the first data beat (\textit{w\_first} or \textit{r\_first}).
    \item \textbf{Data Transfer Time:}  
    From \textit{w\_first} to \textit{w\_last} (writes) or \textit{r\_first} to \textit{r\_last} (reads).
\end{itemize}
If any phase extends beyond its allotted time, the TMU flags a timeout.

\subsection{Prescaler and Sticky Bit Mechanisms}
To further optimize area, both Tc and Fc can reduce the frequency of counter increments through a \textit{prescaler}. Although this lowers resolution, critical events remain detectable. A \textit{sticky bit} ensures that once a near-timeout condition is observed, it remains recorded even if the counter update is delayed by the prescaler.

\subsection{System Observability and Reliability}

By providing real-time tracking of each AXI4 request, the TMU captures latency metrics, identifies bottlenecks, and quickly isolates faulty devices. The Full-Counter solution pinpoints errors down to specific phases, whereas the Tiny-Counter solution offers a cost-effective alternative. In both cases, immediate reset and interrupt capabilities guarantee swift recovery from faults, reinforcing system reliability in safety-critical applications.

\section{Evaluation}\label{AA}

We evaluate the TMU at both the IP level (synthesized standalone) and the system level (integrated into a real SoC).

\subsection{IP-Level Evaluation}
To assess the TMU in isolation, we synthesized it in GlobalFoundries’ 12nm (GF12) technology under four configurations:

\begin{itemize}
    \item \textbf{Tc}: Tiny-Counter without prescaler or sticky bit.
    \item \textbf{Fc}: Full-Counter without prescaler or sticky bit.
    \item \textbf{Tc+Pre}: Tiny-Counter with prescaler and sticky bit.
    \item \textbf{Fc+Pre}: Full-Counter with prescaler and sticky bit.
\end{itemize}

\subsubsection{Outstanding Transaction Setup}
We varied the total number of outstanding transactions from 1 to 128 while fixing the number of unique IDs to 4. Within each ID, we explored between 1 and 32 outstanding transactions. This reflects a typical SoC scenario with multiple masters issuing transfers simultaneously, covering both moderate (e.g., 8-32 total transactions) and high concurrency (e.g., up to 128). Each configuration also supports transactions lasting up to 256 clock cycles to accommodate large bursts.

\subsubsection{Area Results}

\begin{figure}[ht]
    \centering
    \includegraphics[width=0.8\columnwidth,height=4cm]{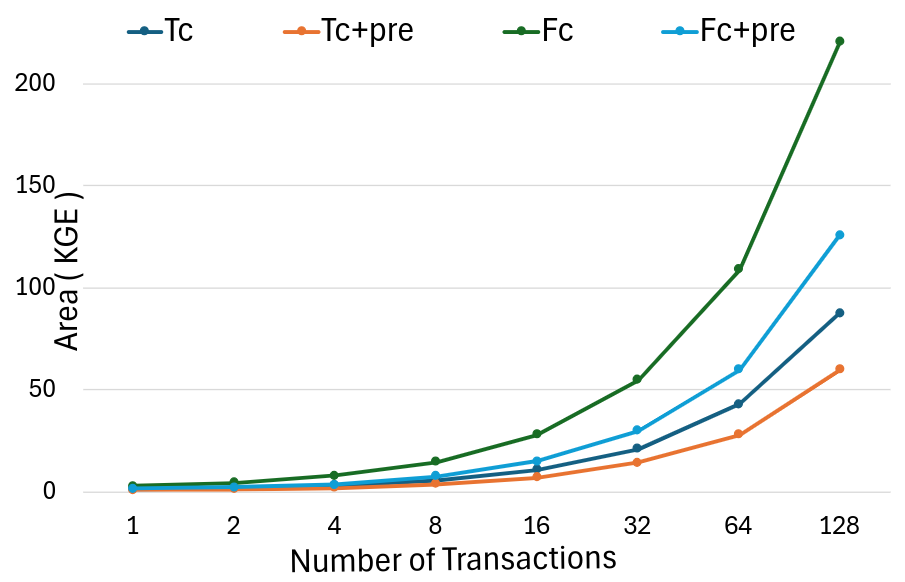}
    \caption{Area comparison of the four configurations (Fc and Tc, with and without prescaler).}
    \label{fig:combined}
\end{figure}

Figure~\ref{fig:combined} shows the area results for all four TMU configurations, both with and without a prescaler, illustrating how area scales with the number of outstanding transactions for Tc and Fc. With a prescaler step of 32, the Tiny-Counter plus prescaler (Tc+Pre) consistently consumes the least area, while the Full-Counter without a prescaler (Fc) is the largest. On average, Tc requires about 38\% of Fc’s area. Across the explored range, prescalers reduce area by 18--39\% (Tc) and 19--32\% (Fc), trading off timing resolution for hardware savings.

\begin{figure}[ht]
    \centering
    \subfloat[Full-Counter Prescaler Exploration\label{fig:presc_fc}]{%
        \includegraphics[width=0.45\columnwidth,height=2.7cm]{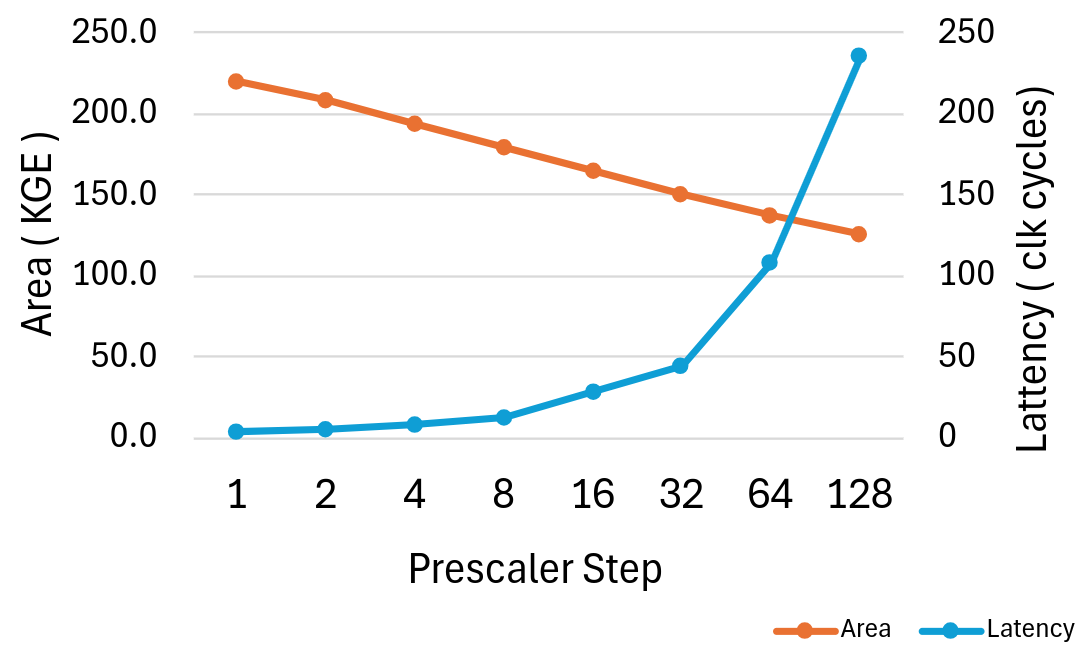}
    }
    \hspace{-0.001\columnwidth}
    \subfloat[Tiny-Counter Prescaler Exploration\label{fig:presc_tc}]{%
        \includegraphics[width=0.45\columnwidth,height=2.7cm]{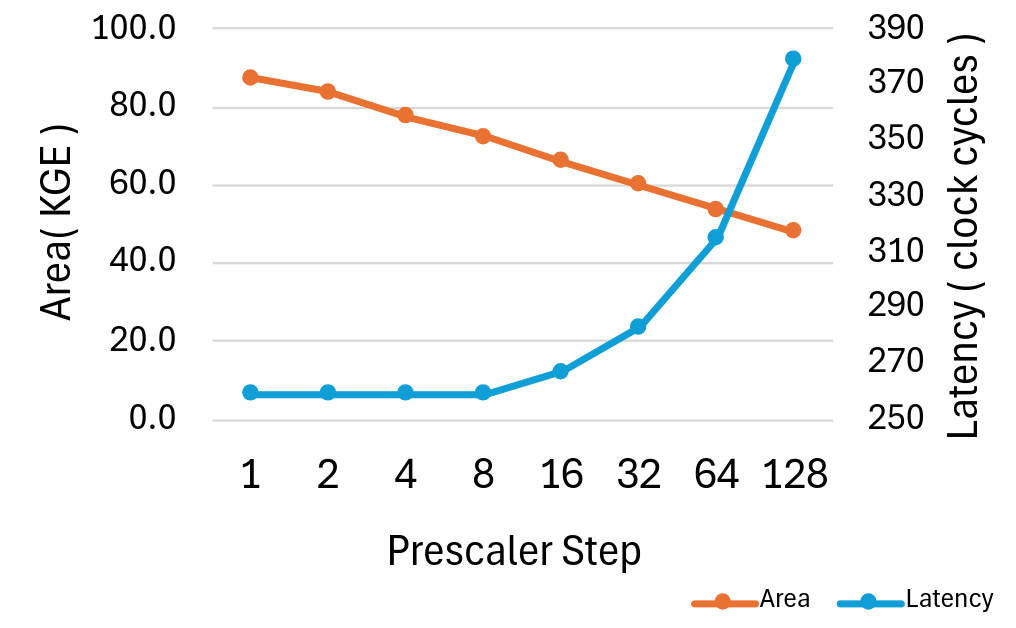}
    }
    \caption{Effect of varying prescaler steps on area and detection latency for Fc (a) and Tc (b), at a fixed 128-outstanding-transaction capacity. Larger prescaler values reduce area but increase fault-detection latency.}
    \label{fig:prescaler}
\end{figure}

To further examine the impact of prescaler settings, Figures~\ref{fig:prescaler}(a) and \ref{fig:prescaler}(b) plot area versus detection latency for Fc and Tc, respectively, at a fixed capacity of 128 outstanding transactions. Here, the prescaler step ranges from 1 to 128. As expected, larger prescaler values reduce area but increase detection latency, since counter updates occur less frequently and thus can be tracked with a smaller counter width. In this setup, the latency is measured under a scenario where the datapath never asserts a valid signal, effectively modeling a total stall condition.

\subsubsection{Fault Injection Tests}
\begin{figure}[htbp]
    \centerline{\includegraphics[width=0.4\textwidth,height=0.2\textwidth]{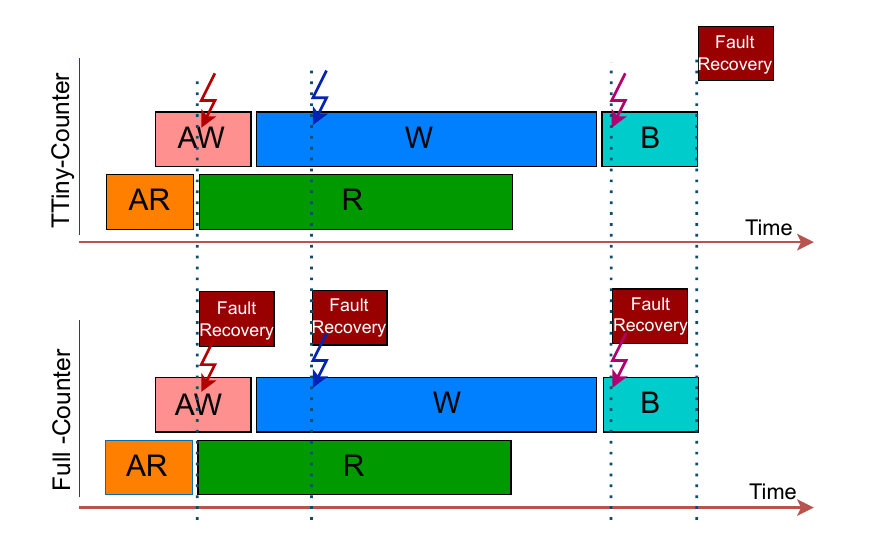}}
    \caption{Fault injection setup for IP-level tests.}
    \label{fig:fault_inj}
\end{figure}

We validated fault detection and latency by injecting random failures at key AXI transaction stages (Fig.~\ref{fig:fault_inj}). These include:
\begin{itemize}
    \item AW Stage Error: Missing \textit{aw\_ready} acknowledgment
    \item W Stage Timeout: No valid data received from the master
    \item W Datapath Error: \textit{w\_ready} failure during data transfer
    \item Data Transfer Error: Issues between \textit{w\_first} and \textit{w\_last}
    \item \textit{w\_last} to \textit{b\_valid} Error
    \item B Handshake Error: Handshake failure or ID mismatch on the B channel
\end{itemize}

Phase-specific counters in the Fc solution detect errors earlier and provide detailed performance logging, but at higher area cost. In contrast, the Tc approach offers a single transaction-level counter that reduces hardware overhead but detects errors only after the full transaction time budget.

Overall, the IP-level results confirm that the TMU scales effectively with varying numbers of outstanding transactions and can be tuned via prescalers to meet different area and performance requirements.

\subsection{System-Level Evaluation}
\begin{figure}[htbp]
    \centerline{\includegraphics[width=0.42\textwidth,height=0.28\textwidth ]{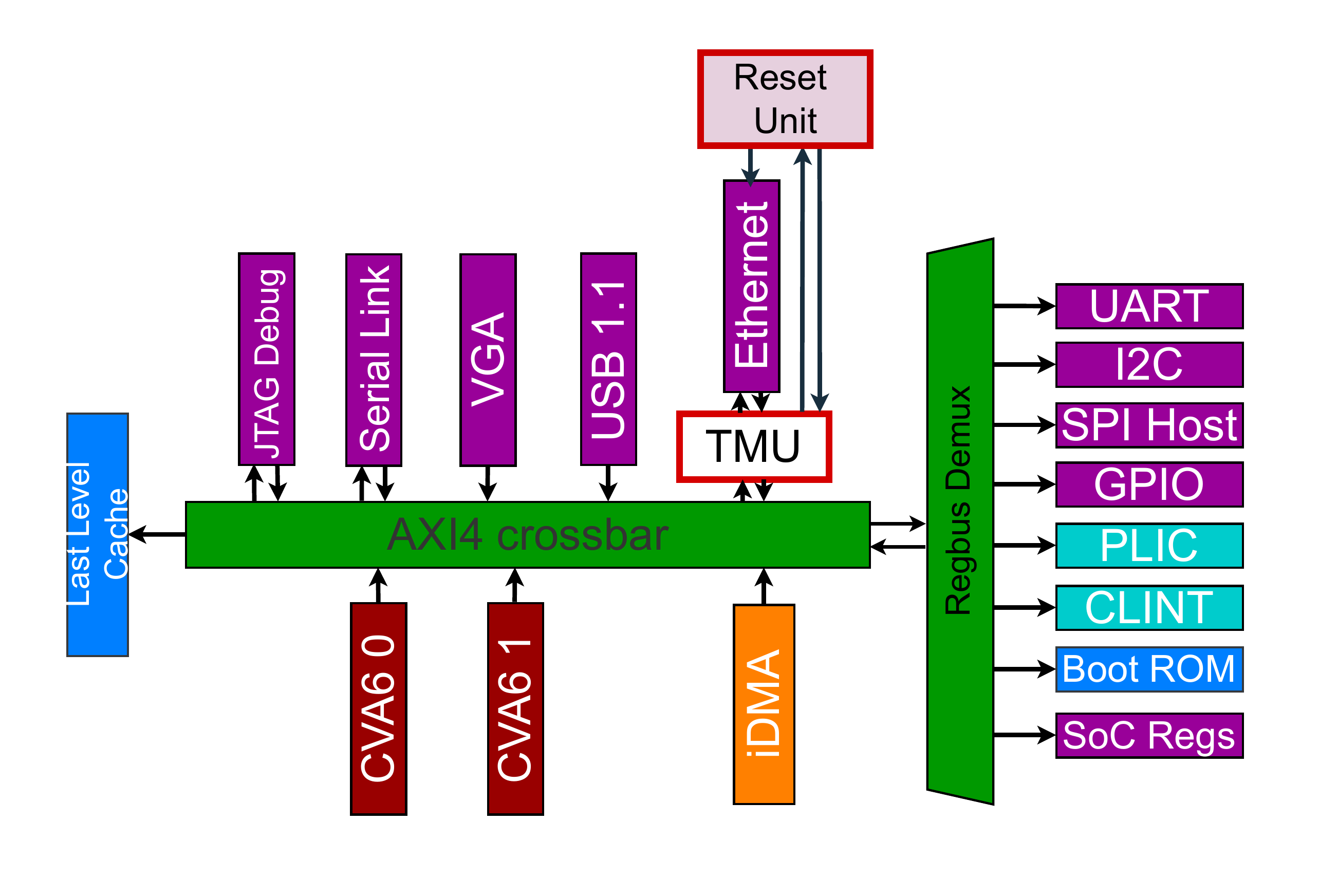}}
    \caption{System-level integration: TMU embedded in Cheshire to monitor Ethernet traffic.}
    \label{fig:cheshire}
\end{figure}

To demonstrate full-system reliability, we integrated the TMU into the Cheshire platform~\cite{b14}, a Linux-capable RISC-V CVA6-based SoC, monitoring an RGMII Ethernet peripheral. As shown in Fig.~\ref{fig:cheshire}, the TMU sits between the AXI crossbar and the Ethernet IP, observing all traffic flowing through Ethernet.

We evaluated a transaction with 250 beats on a 64-bit bus, effectively stressing the Ethernet interface beyond typical packet sizes. The Tiny-Counter (Tc) used a single time budget of 320 clock cycles for the entire transaction, whereas the Full-Counter (Fc) allocated distinct budgets for each phase (e.g., 10 cycles for AW, 250 for W, etc.). Fault injections were identical to those at the IP level. On detecting a timeout or protocol violation, the TMU raises an interrupt and requests an external reset of the Ethernet IP. Upon reset completion, the TMU resumes normal monitoring to ensure continued system stability.

\begin{figure}[htbp]
    \centerline{\includegraphics[width=0.52\textwidth,height=0.25\textwidth]{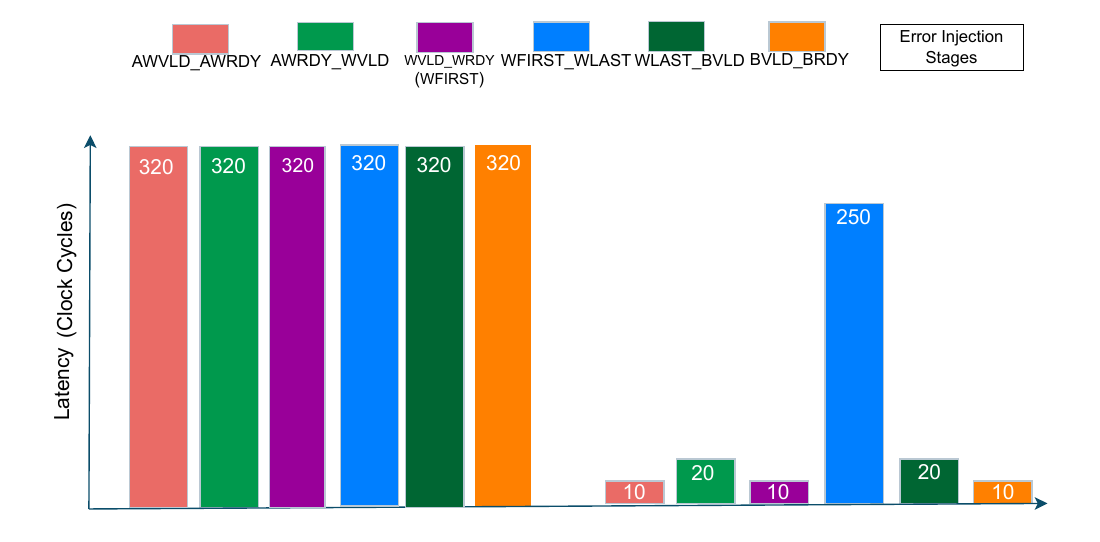}}
    \caption{Latency comparison for Ethernet transactions under different fault injections. Fc detects errors earlier by monitoring each phase, while Tc reports errors only at transaction completion.}
    \label{fig:eth_results}
\end{figure}

Figure~\ref{fig:eth_results} compares detection latencies when injecting errors at the beginning, middle, and end of the transaction. With Tc, detection always occurs after the entire time budget (320 cycles). In contrast, Fc signals a fault once the relevant phase times out, providing near-immediate detection if the error occurs early (e.g., during AW).

These system-level findings show that the TMU effectively isolates and recovers from errors with minimal disruption. The Fc variant offers tighter fault detection at increased area cost, whereas Tc provides area savings at the expense of coarser detection granularity.

\section{Related Work} 
\label{sec:related}

Transaction monitoring in System-on-Chip (SoC) architectures has attracted incresing attention in the contexts of security, performance analysis, fault detection, and reliability. Transaction Monitoring Units (TMUs) are crucial for diagnosing faults in these complex systems, as they monitor transaction events and analyze resource utilization.

\begin{table*}[ht]
\centering
\caption{Comparison of AXI Transaction Monitors in the Literature}
\label{tab:my-table}
\begin{threeparttable}
\centering
\begin{tabular}{|l|c|c|c|c|c|c|c|c|c|c|}
\hline
\textbf{Reference} & \textbf{\begin{tabular}[c]{@{}c@{}}Target \\ Prot.\end{tabular}} & \textbf{\begin{tabular}[c]{@{}c@{}}HW/SW-\\ Based?\end{tabular}} & \textbf{\begin{tabular}[c]{@{}l@{}}Timing\\ Metrics\end{tabular}} & \textbf{\begin{tabular}[c]{@{}c@{}}Transac.\\Level\end{tabular}} & \textbf{\begin{tabular}[c]{@{}c@{}}Phase\\Level\end{tabular}} & \textbf{\begin{tabular}[c]{@{}c@{}}Prot\\ Check\end{tabular}} & \textbf{\begin{tabular}[c]{@{}c@{}}Perf.\\ Metrics\end{tabular}} & \textbf{\begin{tabular}[c]{@{}c@{}}Fault\\ Detection\end{tabular}} & \textbf{\begin{tabular}[c]{@{}c@{}}M.O\\ Supp.\tnote{*}\end{tabular}} & \textbf{Scalab.}\\ 
\hline
Xilinx AXI Timeout \cite{b2} & AXI & HW & \checkmark & \checkmark & \(\times\) & \(\times\) & \(\times\) & \checkmark & \(\times\) & \(\times\) \\ \hline
ARM Watchdog \cite{b3} & APB & HW & \checkmark & \checkmark & \(\times\) & \(\times\) & \(\times\) & \checkmark & \(\times\) & \(\times\) \\ \hline
AMD Perf. Mon. \cite{b4} & AXI & HW & \checkmark & \checkmark & \(\times\) & \(\times\) & \checkmark & \(\times\) & \(\times\) & \(\times\) \\ \hline
Synopsys Smart Mon. \cite{b5} & AXI & HW & \checkmark & \checkmark & \(\times\) & \(\times\) & \checkmark & \(\times\) & \(\times\) & \(\times\) \\ \hline
Lazaro AXI Firewall \cite{b6} & AXI & HW & \(\times\) & \checkmark & \(\times\) & \(\times\) & \(\times\) & \(\times\) & \(\times\) & \(\times\) \\ \hline
Ravi Bus Monitor \cite{b7} & AXI & HW & \checkmark & \checkmark & \(\times\) & \(\times\) & \checkmark & \(\times\) & \(\times\) & \(\times\) \\ \hline
Lee Bus Monitor \cite{b8} & AXI & HW & \checkmark & \checkmark & \(\times\) & \checkmark & \checkmark & \(\times\) & \(\times\) & \(\times\) \\ \hline
Kyung Perf. Mon.\cite{b9} & AXI & HW & \checkmark & \checkmark & \(\times\) & \(\times\) & \checkmark & \(\times\) & \(\times\) & \(\times\) \\ \hline
Chen AXIChecker \cite{b10} & AXI & HW & \(\times\) & \checkmark & \(\times\) & \checkmark & \(\times\) & \(\times\) & \(\times\) & \(\times\) \\ \hline
Tan Perf. Mon. \cite{b11} & AXI & HW & \checkmark & \checkmark & \(\times\) & \(\times\) & \checkmark & \(\times\) & \(\times\) & \(\times\) \\ \hline
Edelman Transac. Mon.\cite{b12} & AXI & SW & \(\times\) & \(\times\) & \checkmark & \(\times\) & \(\times\) & \(\times\) & \(\times\) & \(\times\) \\ \hline
\textbf{This work: Tiny-Counter} & AXI & HW & \checkmark & \checkmark & \(\times\) & \checkmark & \checkmark & \checkmark & \checkmark & \checkmark \\ \hline
\textbf{This work: Full-Counter} & AXI & HW & \checkmark & \(\times\) & \checkmark & \checkmark & \checkmark & \checkmark & \checkmark & \checkmark \\ \hline
\end{tabular}
\begin{tablenotes}
\footnotesize
\item[*] M.O Supp. denotes Multiple Outstanding Support.
\end{tablenotes}
\end{threeparttable}
\end{table*}

Xilinx provides an \emph{AXI Timeout} IP \cite{b2} that detects stalls by tracking the time between address and response phases. If a response exceeds a user-defined window, it flags an error and can issue an interrupt, preventing indefinite waits. However, it focuses primarily on prolonged inactivity and lacks phase-level latency metrics, detailed performance logs, and fine-grained protocol checks. As a result, while it prevents severe bus hangs, it does not deliver the comprehensive fault isolation or performance monitoring required in many safety-critical or mixed-criticality SoC environments.

The ARM SP805 Watchdog \cite{b3} provides fault detection and system protection by safeguarding SoCs against software malfunctions from unresponsive or runaway processes. It generates interrupts or reset signals upon timeout. Synopsys' Smart Monitor \cite{b5} monitors AXI interfaces, capturing AXI bus traffic and computing key performance metrics such as data byte count, throughput, latency, and request/response counts. Similarly, AMD's AXI Performance Monitor \cite{b4} measures significant AXI performance metrics, including bus latency for specific manager/subordinate pairs and memory traffic over defined durations.

A security-focused AXI transaction firewall \cite{b6} protects subordinate devices from malicious managers by filtering transactions based on operation type (write or read), address range, and required bandwidth against predefined rules. While effective in rejecting unauthorized transactions, it does not monitor transaction timing or detect protocol violations, which are essential for reliable systems.

Ravi \cite{b7} introduced a Bus Monitor that captures performance metrics such as transaction count, transfer size, and latency distributions for AXI transactions using hardware counters. Its primary function is to collect behavioral statistics for performance analysis. Lee and Huang \cite{b8} developed a Reconfigurable Bus Monitor Tool Suite for on-chip SoC monitoring, offering protocol checking and performance monitoring of bus transactions. Although it verifies compliance issues like address alignment in burst transactions, it overlooks critical rules such as ID mismatches and out-of-order completions and lacks comprehensive performance statistics and fault detection mechanisms.

Kyung \cite{b9} presented a Performance Monitoring Unit (PMU) with a Bus Monitor (BM) that observes bus transactions and gathers statistics on address ranges, transfer counts, and read/write latency distributions. Configured to divide transactions into three address ranges, it independently counts bus-transaction events using three performance counter sets. However, this BM does not support multiple outstanding transactions and focuses solely on transaction statistics without addressing error handling or protocol violations.

Chen, Ju, and Huang developed AXIChecker \cite{b10}, a rule-based, synthesizable protocol checker for AXI transactions within SoC environments. It enforces 44 protocol rules across manager, subordinate, and default subordinate behaviors to ensure compliance with address alignment, burst transactions, and data ordering. While effective for real-time rule-based transaction verification and logging detected violations for debugging, it lacks performance and timing monitoring, limiting its applicability in performance-critical SoC designs.

Tan et al. \cite{b11} introduced a method for evaluating the performance of AMBA AXI-based SoC designs, focusing on bandwidth and latency calculations. Edelman and Ardeishar \cite{b12} proposed a UVM-based phase-level monitoring approach for simulation-based verification of AXI transactions, capturing transaction phases including address, data, and response within software environments. This method targets pre-silicon verification, ensuring protocol compliance during simulation but does not extend to post-silicon performance monitoring.

Table~\ref{tab:my-table} summarizes the state-of-the-art works in this field. Our TMU differs by offering comprehensive monitoring at both the transaction and phase levels, supporting multiple outstanding transactions, and providing real-time fault detection and recovery. Notably, the TMU includes advanced features like adaptive time budgeting and detailed performance logging, making it suitable for safety-critical AXI4 systems where in-field fault tolerance is essential. Further, its configurability permits mixing Tiny-Counter and Full-Counter monitors within the same SoC, tailoring overhead and detection granularity to each subordinate’s requirements.

\section{Conclusion}

We presented a versatile Transaction Monitoring Unit (TMU) for AXI4-based SoCs in safety-critical domains such as automotive and aerospace. The TMU provides transaction-level and phase-level monitoring, enabling real-time detection of protocol violations and timeouts, along with rapid fault recovery. Two implementations includign a Tiny-Counter for minimal area and a Full-Counter for more detailed, stage-level monitoring which handles multiple outstanding transactions, adapt time budgets, and log performance metrics. Overall, the TMU delivers a robust, drop-in solution for enhancing reliability, scalability, and performance monitoring in complex AXI4-based SoCs.

\section*{Acknowledgment}

This work was supported in part by Spoke 1 on Future High-Performance Computing (HPC) of the Italian Research Center on High-Performance Computing, Big Data, and Quantum Computing (ICSC), which received funding from the Ministry of University and Research (MUR) under Mission 4 of the ``NextGenerationEU'' program.

\clearpage
\twocolumn

\vspace{12pt}

\end{document}